\documentclass[preprint]{aastex}
\slugcomment{Submitted to ApJ}
\shortauthors{Zheng et al.}
\begin{document}

\title{An Extreme Ultraviolet Wave Associated with A {\bf Solar} Filament Activation}
\author{Ruisheng Zheng, Yao Chen, Bing Wang, and Hongqiang Song}
\affil{Shandong Key Laboratory of Optical Astronomy and Solar-Terrestrial Environment, School of Space Science and Physics, Institute of Space Sciences, Shandong University,  Weihai, Shandong, 264209, China; ruishengzheng@sdu.edu.cn}

\begin{abstract}
Extreme ultraviolet (EUV) waves are impressive coronal propagating disturbances. They are closely associated with various eruptions, and can used for the global coronal seismology and the acceleration of solar energetic particles. Hence, the study of EUV waves plays an important role in solar eruptions and Space Weather. Here we present an EUV wave associated with a filament activation that did not evolve into any eruption. Due to the continuous magnetic flux emergence and cancellation around its one end, the filament rose with untwisting motion, and the filament mass flowed towards another end along the rising fields. Intriguingly, following the filament activation, an EUV wave formed with a fast constant speed ($\sim$500 km s$^{-1}$) ahead of the mass flow, and the overlying coronal loops expanded both in lateral and radial directions. Excluding the possibility of a remote flare and  an absent coronal mass ejection, we suggest that the EUV wave was only closely associated with the filament activation. Furthermore, their intimate spacial and temporal relationship indicates that the EUV wave was likely directly triggered by the lateral expansion of overlying loops. We propose that the EUV wave can be interpreted as linear fast-mode wave, and the most vital key for the successful generation of the EUV wave is the impulsive early-phase lateral expansion of overlying loops that was driven by the activated filament mass flow without any eruption.

\end{abstract}


\section{Introduction}
Extreme ultraviolet (EUV) waves are interesting coronal disturbances propagating at a wide temperature range of EUV emission. EUV waves can provide potential diagnostics on the coronal physical information that cannot be measured directly, such as some plasma parameters (e.g., densities, heating functions, and transport coefficients) and the coronal magnetic field strength (Ballai 2007; Ballai \& Douglas 2008; Kwon et al. 2013; Long et al. 2013). EUV waves can also be one possible source of the acceleration of solar energetic particles that have significance for Space Weather (Park et al. 2013; Miteva et al. 2014). Since their discovery, the physical nature of EUV waves has been strongly debated, and a number of models were proposed~\citep{thompson98,chen02,zhukov04,ballai05, attrill07, wang09}. Benefiting from three generations of space-borne EUV telescopes, especially the latest Atmospheric Imaging Assembly (AIA; Lemen et al. 2012) on board the Solar Dynamics Observatory (SDO; Pesnell et al. 2012), EUV waves have been best interpreted as a bimodal composition of an outer fast-mode MHD wave and an inner non-wave component of coronal mass ejections~\citep*[CMEs;][]{liu10,chen11,downs12,liu14}.

As one basic characteristic of EUV waves, kinematics is crucial for their physical nature. EUV waves are usually detected propagating at velocities of several hundred km s$^{-1}$, and can occasionally speed up to 1000-2000 km s$^{-1}$ (Wills-Davey et al. 2007). Most of EUV waves propagate nearly at constant speeds or with decelerations, and a few of examples exhibit accelerations or very erratic kinematics (Zhukov et al. et al. 2009; Liu et al. 2010; Xue et al. 2013; Zheng et al. 2019). Based on the different kinematics, it has been proposed that EUV waves can be interpreted as nonlinear fast-mode waves or shocks with stronger deceleration at higher speeds, linear fast-mode waves with nearly uniform moderate speeds, and magnetic reconfiguration non-waves (Warmuth \& Mann 2011).  However, using the AIA ensemble, Nitta et al. (2013) found there are weak related decelerations for cases with higher speeds of 450-800 km s$^{-1}$, and suggested that the noticeable decelerations could be due to the interactions with coronal structures.

EUV waves are always associated with a variety of energetic eruptions, such as CMEs, flares, and filament eruptions~\citep{nitta13}. It is widely recognized that the onset of EUV waves strongly depends on the lateral expansion of the CME flank and is weakly related to flares~\citep{biesecker02,cliver05,chen06}. Furthermore, there is a noticeable trend that the impulsiveness of the early-phase CME lateral expansion is the most crucial key for the EUV wave generation, regardless of the associated CME final speeds or flare sizes (Patsourakos \& Vourlidas 2012; Liu \& Ofman 2014). Eruptions without considerable (or with non-impulsive) lateral expansion trigger no or weak EUV wave. In addition, EUV waves can also be driven by small-scale jet-like eruptions, even though the eruption is failed to form CMEs~\citep{zheng12a,zheng12b,shen18}. Comparing with jet-driven waves and flux-rope-driven ones, ~\cite{zheng19} suggested that EUV waves are likely triggered by the sudden expansion of the closed loops ahead of the ejecting jets or erupting flux ropes. More details about EUV waves can be found in recent reviews~\citep{gallagher11,patsourakos12,liu14,warmuth15,chen16,long17}.

However, it has never been reported for an EUV wave associated without any eruption. Here we report an EUV wave only associated with an filament activation that did not form any eruption on 2015 May 7, combining with observations from AIA and Helioseismic and Magnetic Imager (HMI; Scherrer et al. 2012) onboard SDO, and from Global Oscillation Network Group (GONG) of the National Solar Observatory. The event sheds light on the true driving mechanism of EUV waves.

\section{Observations}
We mainly employe the observations from the AIA on SDO, and the event was visible in all AIA EUV channels that contain a wide range of temperatures. The AIA image (4096~$\times$ 4096 pixels) covers the solar full disk and up to 0.5 $R_\odot$ above the limb, with a pixel resolution of 0.6$"$ and a cadence of 12 s. To identify the associated coronal response at high altitude, we check the EUV images with an extended field of view (FOV) of 54$'$, from the Sun Watcher using Active Pixel System detector and Image Processing (SWAP; Halain et al. 2013; Seaton et al. 2013) of the Project for On Board Autonomy 2 (PROBA2; Santandrea et al. 2013).

Magnetograms from HMI on SDO, with a cadence of 45 s and pixel scale of 0.6$"$, are also used to check the magnetic field configuration of the source region. The filaments are identified by H$\alpha$ filtergrams at 6563~{\AA} from GONG, with a spatial resolution of 1$"$ and a cadence of $\sim$1 minute. On the other hand, we also capture the faint trajectories of the activated filament mass by the technique of persistence mapping (Thompson \& Young 2016), in which the value at each pixel is chosen as the maximum one at this same pixel among a set of images. The coronal magnetic filed lines are extrapolated with the Potential Field Source Surface (PFSS; Schrijver {\&} De Rosa 2003) package of SolarSoftWare. The kinematics of the EUV wave, the loop expansion, and the filament mass flow, are obtained by the time-slice approach.

\section{Results}
\subsection{Filament and Magnetic Activities}
The overview of the filaments before the activation is shown in Figure 1. In H$\alpha$ (panel (b)), the activated filament was the short one (F1), and two more clumps of dark filament mass (F2a and F2b) were fragments of another long filament (F2; dotted lines connecting F2a and F2b). In AIA EUV images (panels (d)-(f)), F1 became bright, and F2 was still dark (green arrows). Checking with the PFSS extrapolation method, it is clear that the extrapolated lower field lines (yellow lines) just overlay F1 and F2, and connected the dispersed negative and positive polarities (the red minus and plus) in the north edge of active region (AR) 12339  (panels (a)-(b)). The extrapolated lower filed lines were also consistent with the coronal loops in SWAP 174~{\AA} and AIA 171~{\AA} images (panels (c) and (e)). Hence, F1 and F2 were in the same filament channel. Superimposing the filaments (green contours) on the HMI magnetogram (panel (a)), the south end of F1 clearly anchored at the north boundary region of AR 12339, as the source region  (blue boxes in panels (a)-(b) and (d)) of the following filament activation.

The source region was dominated by two small negative polarities (in the enlarged view in the left upper corner of Figure 1(a)), and the magnetic field evolution in the source region is displayed in Figure 2. It is obvious that, on the east of the north negative polarity, a minor positive polarity quickly emerged and fast disappeared in two hours (black arrows in panels (a)-(c)). It is possible that there existed the magnetic flux cancellation between the emerging positive polarity and the nearby north negative polarity around the south end of F1. The evolution of total unsigned magnetic flux in the region (panel (d)) clearly shows a continuous emergence and cancellation of magnetic flux, and the beginning of the filament activation (the vertical dashed line) is just during the rapid decrease of the magnetic flux. It implies a close relationship between the continuous magnetic activities and the filament activation.

\subsection{Filament Activation}
The filament activation is shown in Figure 3 (also see associated animation). In H$\alpha$ filtergrams (panels (a) and (d)), due to the activation, F1 slowly disappeared, but the mass flow was faint. In AIA 304 and 171~{\AA} images, the mass flow of F1 was in the form of three successive bright jets (green arrows in panels (b)-(c) and (g)), and the trajectories delineated loop-like magnetic structures during the ascent (green arrows in panels (e)-(f)). The activated mass fell down between F2a and F2b, which resulted in some brightenings beneath F2 (blue arrows in panel (f)) and the horizontal mass flow of F2 in the filament channel. Note that there existed a weak untwisting motion for the rising F1. Along the direction perpendicular to the south end of the rising F1 (S1; the dotted line with a triangle start in panel (c)), the time-distance intensity plot (panel (h)) shows that only the first and second jets had a weak untwisting motion with a speed of $\sim$6 km s$^{-1}$ (green and blue dotted lines), and the third jet did not present the untwisting motion. Along the direction perpendicular to the top of the rising F1 (S2; the dotted line with a triangle start in panel (e)), the time-distance intensity plot (panel (i)) shows that the ascent speed of the activated F1 was $\sim$6 km s$^{-1}$ between the first and second jets, and then decreased to $\sim$2 km s$^{-1}$ until the third jet (blue and black dotted lines). The time-distance plots show the rising and untwisting motion was mainly associated with the first and second jets.

The activated F1 and F2 are much obvious in negative persistence images between 19:20 and 20:40 UT in AIA 304, 171 and 193~{\AA} (Figure 4). The three loop-like structures (colored dotted lines) indicate three paths (P1, P2, P3) of the rising F1 in the chronological order with increasing heights (panels (a)-(c)). In the time-distance negative plots in AIA 304~{\AA} along P1-P3 (panels (d)-(f)), it is clear that the activated mass flow dominated in a sequential heights from P1 to P3, and the discontinuous mass along P1 and P2 likely implies the untwisting motion of the rising F1 (black arrows). The flow speeds along P1-P2 ($\sim$150-170 km s$^{-1}$) are less than that along P3 ($\sim$200 km s$^{-1}$), which is likely because that the untwisting motion dominated in the early phase (P1-P2) of the filament activation.

\subsection{EUV Wave and Loop Expansion}
Following the filament activation, an EUV wave occurred ahead of the mass flow (Figure 5 and its animation). In running-ratio-difference images in AIA 193~{\AA} (panels (a)-(b)), the wave (green arrows) emanated close to the rising loop-like field (the triangle and the red arrow), and primarily propagated northwestwards. In the composite base- (panel (c)) and running- (panel (d)) ratio-difference images in AIA 171 (red), 193 (green) and 211 (blue)~{\AA}, besides the distinct wavefront (green arrows), coronal dimmings (the white arrow) were also obvious, as the coronal response for the filament activation. The kinematics evolution of the wave front is shown in time-distance plots of the running-ratio-difference (panel (e)) and normalised-intensity-original (panel (f)) along the selected slice (S3; the dotted line in panel (a)). S3 extends northwestwards from the turn of the loop-like fields that is consistent with the position of the weak brightening at $\sim$19:56 UT (green triangles in panels (a) and (e)-(f)). The wave had a fast constant velocity of $\sim$500 km s$^{-1}$ (the green arrow and green dotted line), and only lasted for a few of minutes. Just before the wave onset (the diamond), it is very clear that there existed the lateral expansion of the loops ahead of the filament channel, with a speed of $\sim$110 km s$^{-1}$ (the blue arrow and blue dotted line). Note that dense coronal loops (the black arrow) occupied between the activated F1 (the triangle) and the wave onset (the diamond), and were replaced by coronal dimmings (the white arrow) after the expansion (panel (f)). It is apparent for the intimate temporal and spacial relationship between the wave onset and the lateral expansion of overlying loops.

Apart from the lateral expansion at their flanks, the coronal loops over the filament channel also expanded radially at the top in AIA 131 and 94~{\AA} (Figure 6 and its animation). During the filament activation (the blue arrows in panels (a)-(c)), obvious magnetic rope-like structures surrounding F1-F2 appeared and then fade away later in AIA 94~{\AA} (white arrows in panels (d)-(f)). Intriguingly, a cluster of faint coronal loops came out above the activated filaments and slowly expanded outwards radially some minutes later than the wave onset (yellow arrows in panels (c)-(e)). Finally, the tops of expanded loops reconnected with the nearby open field lines (black arrows), and evolved into newly-formed open fields (the orange arrow in panel (f)). Note that, during the filament activation, there occurred a C5.0 flare with a fan-spine topology in the core of AR 12339 (the red and green arrows) in the persistence image between 19:20 and 20:40 UT in AIA 94~{\AA}, and the flare center was far away from the site of wave onset (the diamond and white arrow in panel (a)). The outer spine (the red arrows in panels (d)-(g)) emanating from the flare center extended eastwards, which was clearly separated from the coronal structures of the filament activation. In the light curve of the flare region (the dotted box in panel (a)) in 94~{\AA} (panel (h)), it is clear that the flare started at $\sim$19:48 UT (the dashed vertical line).

\section{Conclusions and Discussion}
Combining with the PFSS extrapolated field lines, HMI magnetograms, AIA images, and H$\alpha$ filtergrams, F1 and F2 obviously lay in the same filament channel, covered by a series of lower coronal loops (Figure 1). Around the south end of F1, there occurred the rapid emergence and cancellation of magnetic flux around the south end of F1 (Figure 2), which likely resulted in the activation of F1. As a result, the mass at the south end of activated F1 transferred to another end in the form of jets, and finally fell down beneath F2 in the same filament channel (Figure 3(a)-(g)), which means the filament activation did not form any eruption. The mass flow along the F1 fields displayed distinct rising and untwisting motion during the first and second jets (Figure 3(h)-(i) and Figure 4), and the unwinding motion speed ($\sim$6 km s$^{-1}$) is similar to those in similar cases of filament activation (Yang et al. 2018; Zheng et al. 2018). It is possibly resulted from the magnetic reconnection between the emerging positive magnetic flux and part of F1 twisted fields. Intriguingly, following the filament activation, an EUV wave formed ahead of the mass flow, and the overlying loops clearly expanded both in lateral and radial directions (Figure 5 and 6). Moreover, the rope-like structures at high temperature in AIA 94~{\AA} (white arrows in Figure 6 (d)-(f)) is probably heated by the released energy from the activation of the inner filament (F1).

Interestingly, the EUV wave had a fast constant speed of $\sim$500 km s$^{-1}$. Because the propagation is mostly perpendicular to the magnetic field lines of expanding loops, the EUV wave with a constant speed can be treated as the linear fast-mode wave. The characteristic velocity of the fast-mode wave is calculated by
$v_f = \sqrt{v_{A}^2 + c_{s}^2}$ , where $v_A = B/{\sqrt{4 \pi m_{p}n}}$ is the Alfv{\'e}n speed, and $c_{s}$ is the sound speed. $B$ is the radial component of the nearby coronal magnetic field. As this EUV wave mainly propagated in the quiet low corona, it is set for the electron number density of $10^8$ cm$^{-3}$, and a typical coronal sound speed of 180 km s$^{-1}$. With the fast-mode wave speed of 500 km s$^{-1}$, the magnetic field strength can be derived as $B = 2.15$ G, which is in the typical range of $B = 2-6$ G for the quiet corona (Long et al. 2013). Hence, It is reliable for the fast constant velocity of the EUV wave.

In the LASCO CME catalog, there are only two CMEs{\footnote{\url{https://cdaw.gsfc.nasa.gov/CME\_list/UNIVERSAL/2015\_05/univ2015\_05.html}}} emanating from the quadrant over the region of filament activation. However, these CMEs happened hours before the filament activation, and launched from a different AR (the black arrow in Figure 1(c)). Hence, the filament activation did not form any CME that can be associate with the EUV wave. On the other hand, there occurred a C5.0 flare in the core of AR 12339 at $\sim$19:48 UT (Figure 6(h)), before the wave onset ($\sim$19:58 UT; Figure 5(e)-(f)) but after the beginning of the filament activation ($\sim$19:28 UT; Figure 2(d)). As the case of M2.0 flare in Zheng et al. (2019), the eruption in the fan-spine topology likely triggered an EUV wave ahead of the rapid expansion of the coronal loops of the outer spine that represents the direction of eruption and energy release. Here, the post-flare outer spine clearly extended in an orthogonal direction (red arrows in Figure 6) relative to that of the filament activation and the wave propagation. It likely indicates the fan-spine structure in the AR core was on one side of filament channel that rooted in the AR north edge. The flare can disturb the filed lines of filament channel to reduce the magnetic pressure above filaments ({\bf Jenkins et al. 2018, 2019}), but is impossible to invoke the rapid lateral expansion of the overlying filament loops and a following narrow EUV wave. It is suggested that the flare was barely associated with the EUV wave. Therefore, the EUV wave was only closely associated with the filament activation that was in the form of mass flow with rising and untwisting motion.

How was the EUV wave related to the filament activation? The filament channel involving F1 and F2 was covered by lower coronal loops (the extrapolated field lines in Figure 1). The filament activation forced the overlying loops to expand in the direction of the filament mass flow, which brought the coronal dimmings and the lateral loop expansion (the white and blue arrows in Figure 5). The time-distance plots show clearly that the waves have a close spacial-temporal relationship with the laterally expanding loops (panels (e)-(f) in Figure 5). Hence, it is believed that the wave was likely triggered by the lateral loop expansion pushed by the filament activation below, which is similar to the trigger of lateral expansion of loops over an erupting core (flux ropes or jets). The filament activation was much weaker than any eruption with energetic erupting core, but the mass flow with untwisting motion may contribute to the lateral loop expansion.

How did the slow mass flow and loop expansion ($\sim$150 and $\sim$110 km s$^{-1}$) trigger a fast EUV wave ($\sim$500 km s$^{-1}$)? It has been widely recognised that the impulsive early-phase lateral expansion is crucial for the EUV wave generation, regardless of the final speeds of associated CMEs, and eruptions with non-impulsive lateral expansion cannot produce any EUV wave (Liu \& Ofman 2014). In this case, the well-defined wave signature reveals the occurrence of the impulsive lateral expansion at the beginning, and the initial lateral expansion speed is likely close to the fast speed of EUV wave. However, there are dense coronal loops beyond the activated filament (the black arrow in Figure 5(f)). During the propagation through the dense loops, the linear fast-mode wave keeps the constant speed of $\sim$500 km s$^{-1}$, but the lateral expansion rapidly decreased to $\sim$110 km s$^{-1}$, similar to the separation between CME flanks and EUV wave fronts. Hence, the speed difference here between the lateral expansion and EUV wave should be at the late stage.

Therefore, we suggest a scenario of the generation of the EUV wave associated with filament activation in side-view (left panels) and top-view (right panels) of the schematic representation (Figure 7). Before the activation (top panels), F1 (blue) and F2 (cyan) lay in the same filament channel that was covered by the overlying loops (yellow). Due to the emergence of magnetic flux near the northeast end of F1, part of filament fields took part in magnetic reconnections, and F1 was activated. As a result (middle panels), the reconnected fields rose with untwisting motion, and the filament mass nearby began to flow to the other end along the filament fields. Though the mass of F1 finally fell down beneath F2, the horizontal mass flow and untwisting motion forced the overlying loops to impulsively expand in lateral and radial directions. Sequentially, the rapid lateral expansion gave a birth of the EUV wave (the bright yellow shade) at a high initial speed, just ahead of the laterally expanding loops. During the wave propagation (bottom panels), the lateral expansion of overlying loops slowed down to a much slow speed, and the radial expansion was predominant. Therefore, the linear fast-mode EUV wave travelled outward at the constant speed, and left the expanding loops far behind.

All the results showed that the EUV wave closely associated with the filament activation that did not form any eruption, and was likely directly triggered by the sudden lateral expansion of overlying loops above the filament channel. Therefore, we suggest that the EUV wave can form in any kind of solar activity, even no associated eruption, and the impulsive lateral loop expansion at the early stage is the most important condition for the true trigger mechanism. There should exist much more hidden waves associated with non-eruption activities, and they can complement the understanding on their physical nature and trigger mechanism. Further more and better observations are necessary to verify the suggestions.

\acknowledgments
SDO is a mission of NASA's Living With a Star Program. We gratefully acknowledge the usage of data from the SDO, SWAP, and from the ground-based GONG project. This work is supported by grants NSFC 11790303 and U1731101, and Young Scholars Program of Shandong University, Weihai, 2016WHWLJH07. H. Q. Song is supported by the Natural Science Foundation of Shandong Province JQ201710.

\clearpage

\begin{figure}
\epsscale{0.8} \plotone{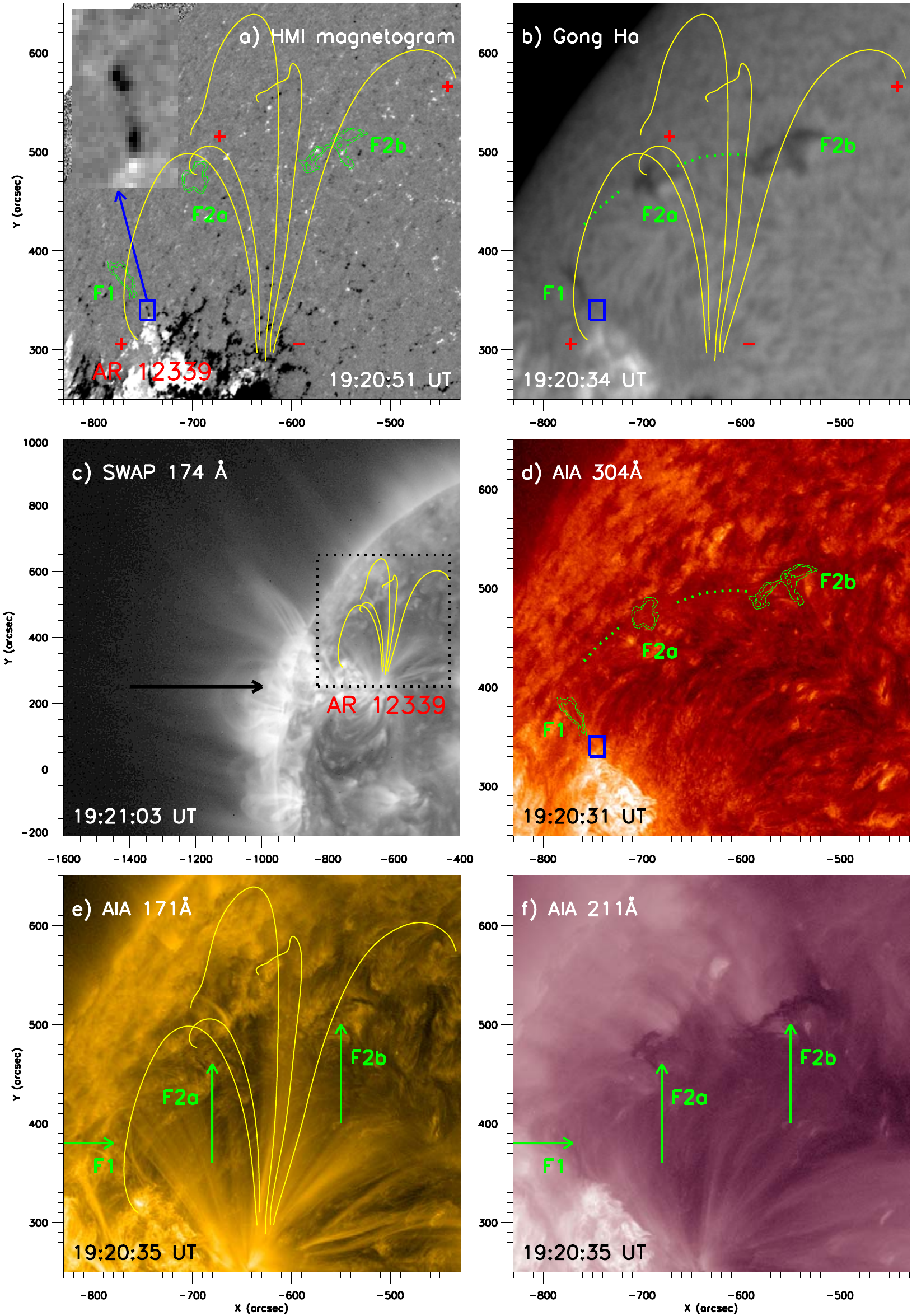}
\caption{The overview of filaments (F1, F2a, F2b; green arrows and contours with dotted lines) before the activation in HMI magnetogram (a), GONG H$\alpha$ filtergram (b), SWAP 174~{\AA} (c), and AIA 304, 171, and 211~{\AA} (d)-(f). The yellow lines are the extrapolated PFSS field lines, and their footpoint polarities are marked with the minus and the plus. The blue boxes indicate the source region of filament activation, and is shown in an enlarged view in the upper left corner in panel (a). The black box in panel (c) represents the FOV of other panels, and the black arrow shows the source of an irrelevant eruption.
\label{f1}}
\end{figure}

\clearpage

\begin{figure}
\epsscale{0.9} \plotone{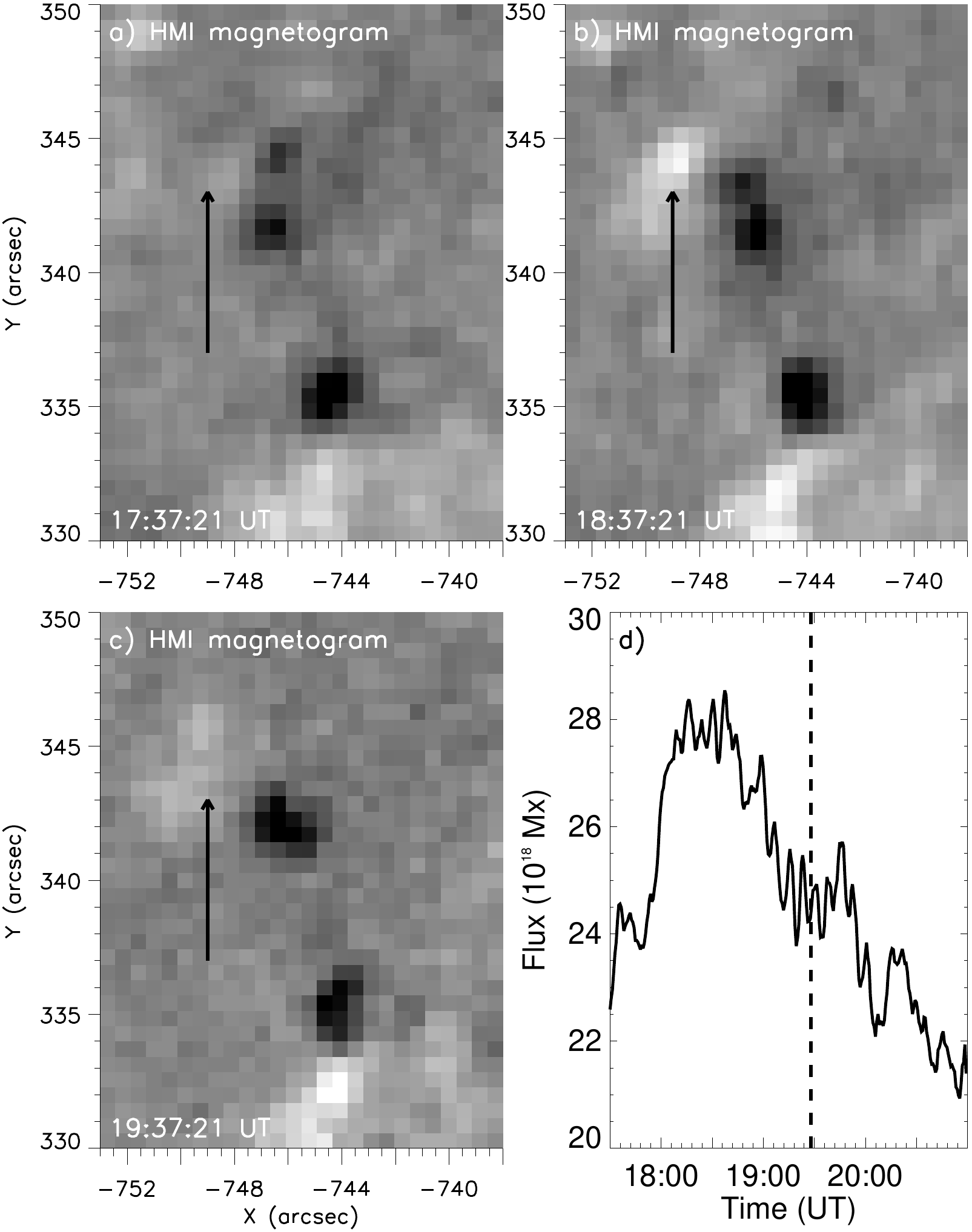}
\caption{The magnetic filed evolution of the source region of filament activation. (a-c) HMI magnetograms showing the growth and disappearance of a minor positive polarity (black arrows), with saturation levels of $\pm$ 100 G. The FOV is indicated by blue boxes in Figure 1. (d) The change of total unsigned magnetic flux in the source region. The vertical dashed line represents the beginning of the filament activation.
\label{f2}}
\end{figure}

\clearpage

\begin{figure}
\epsscale{0.9} \plotone{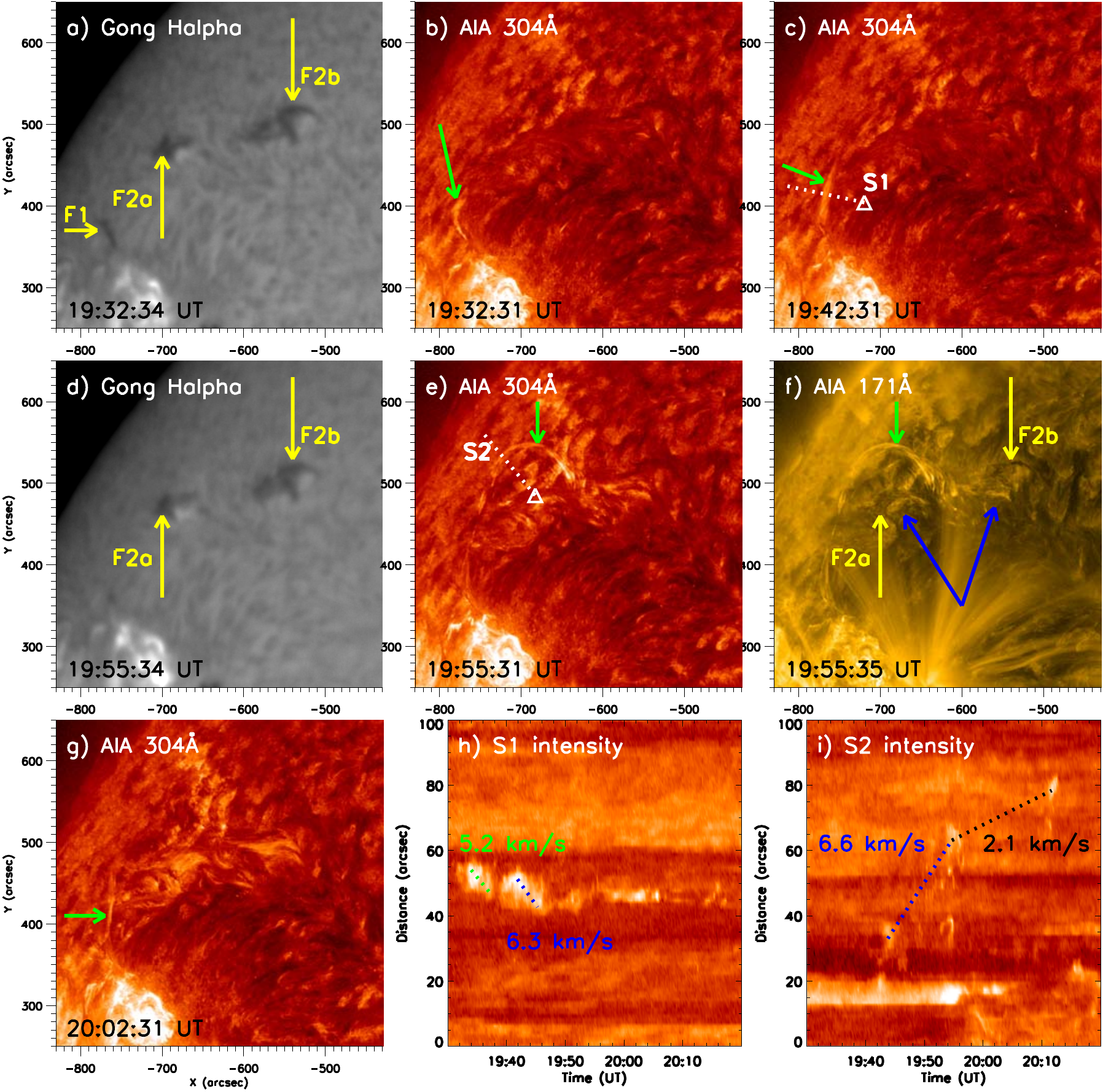}
\caption{The filament activation. (a-g) The activated mass flow of F1 in the form of jets (green arrows) in H$\alpha$ filtergram, and AIA 304 and 171~{\AA}. The yellow arrows indicate F1 and F2, and the blue arrows show the brightenings beneath F2. (h-i) Time-distance intensity plots along S1 and S2 (dotted lines with a triangle start in panels (c) and (e)) in AIA 304~{\AA}. The dotted lines are used to derive the attached speeds.
An animation of the H$\alpha$ filtergram, and AIA 304 and 171~{\AA} images is available. The video begins around 19:20:33 UT and ends at approximately at 20:09:30 UT. The realtime duration of the video is 2 seconds.
\label{f3}}
\end{figure}

\clearpage

\begin{figure}
\epsscale{0.8}
\plotone{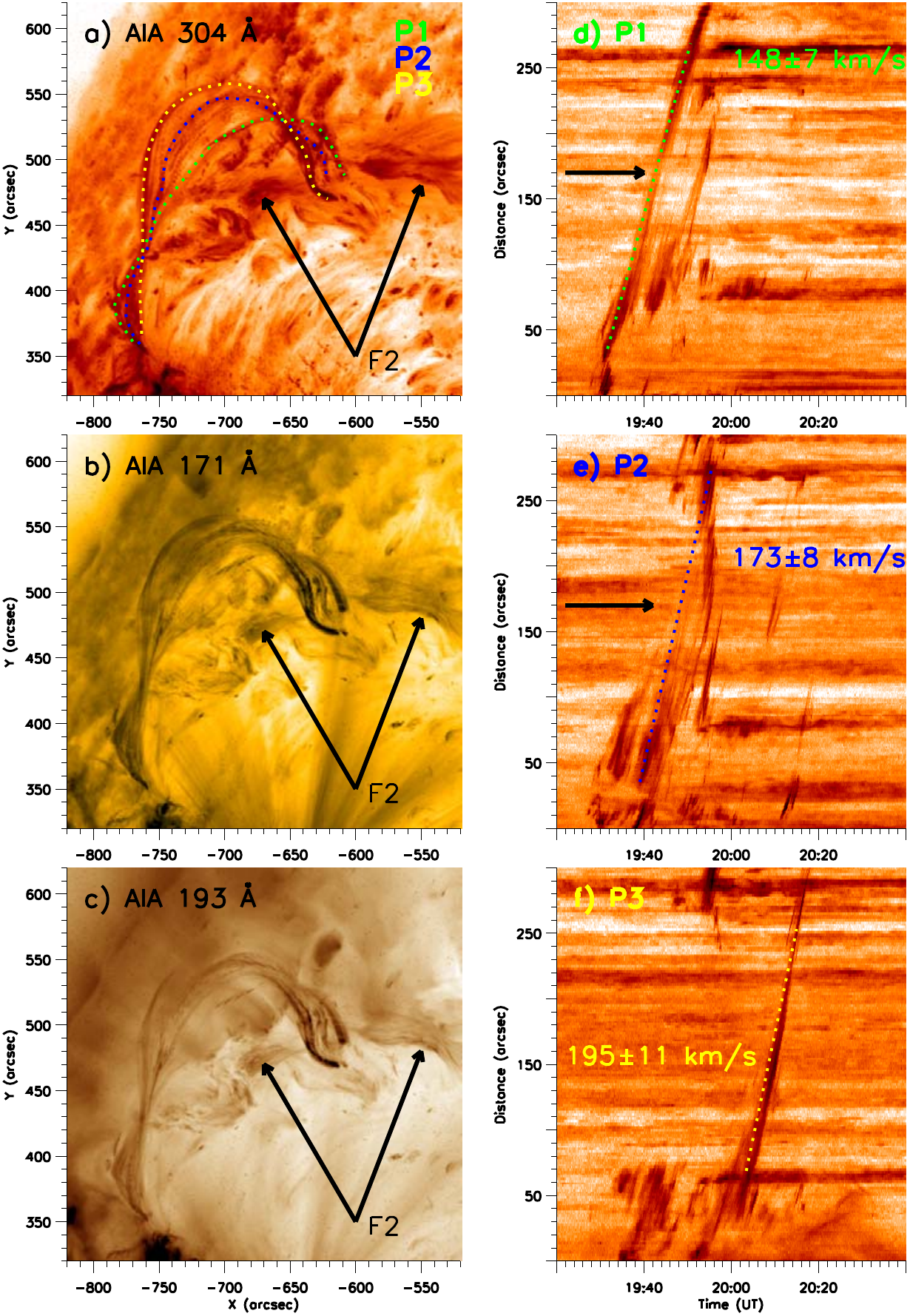}
\caption{The rising and untwisting of activated filament. (a-c) The negative persistence images of AIA 304, 171, and 193~{\AA} between 19:20 and 20:40 UT, outline F1 at three different path (P1-P3; colorful dotted curved lines) and F2 (black arrows). (d-f) Time-distance plots along P1-P3 of negative images in AIA 304~{\AA}, show the mass flow at different times. The dotted lines are used to derive the attached speeds, and the black arrows imply the discontinuousness of the mass flow.
\label{f4}}
\end{figure}

\clearpage

\begin{figure}
\epsscale{0.95}
\plotone{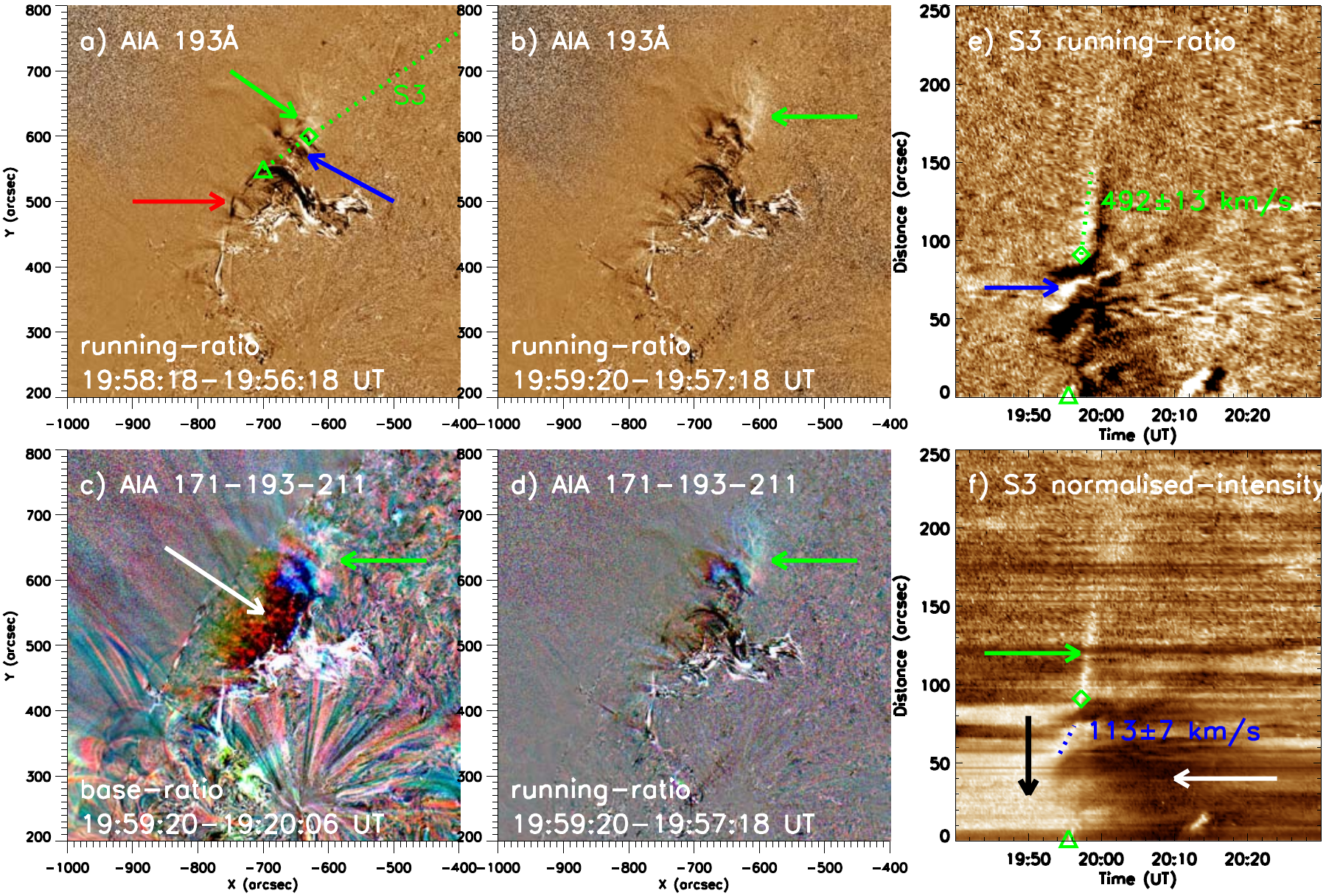}
\caption{The EUV wave. (a-b) Running-ratio-difference AIA 193~{\AA} images showing the EUV wave (green arrows) and the mass flow of F1 (the red arrow). (c-d) Composite base/running-ratio-difference images in AIA 171 (red), 193 (green) and 211 (blue)~{\AA} revealing better the wave (green arrows) and associated coronal dimmings (the white arrow). (e-f) Time-distance plots along S1 in panel (a) of running-ratio-difference and normalised-intensity-original images in AIA 193~{\AA} uncovering the wave propagation, associated dimmings (the white arrow), and the loop expansion (the blue arrow). The triangle and the diamond are represent the start of S1 and the location of the wave onset, and the black arrow shows the dense coronal loops ahead of the activated F1. The dotted lines are used to derive the attached speeds.
An animation of the running-ratio-difference AIA 193~{\AA} plus the composite base/running-ratio-difference images in AIA 171 (red), 193 (green) and 211 (blue)~{\AA} images is available. The realtime duration of the video is 1 second.
\label{f5}}
\end{figure}

\clearpage

\begin{figure}
\epsscale{0.6} \plotone{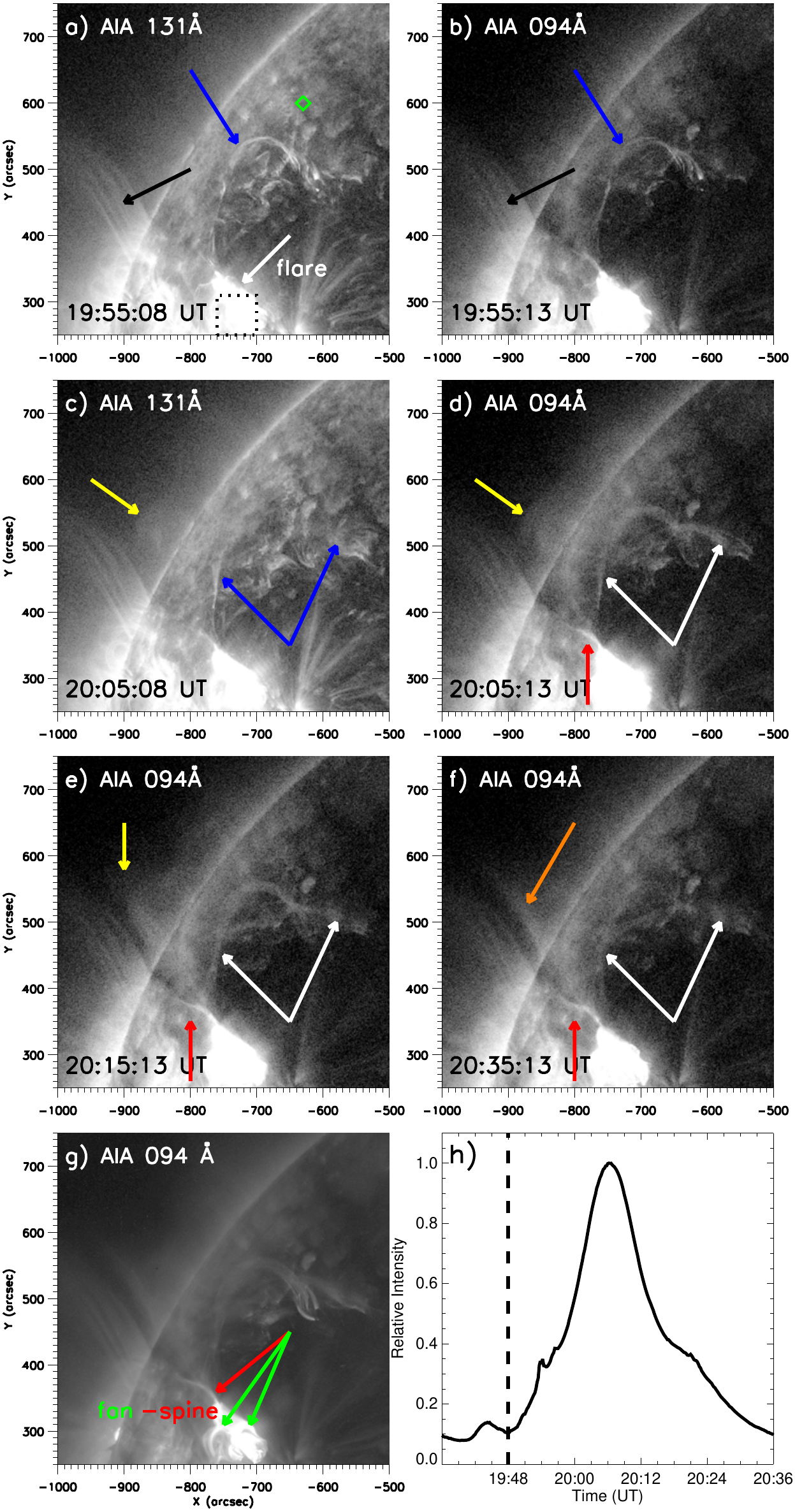}
\caption{The loop radial expansion. (a-f) AIA 131 and 94~{\AA} images showing the radial expansion of loops (yellow arrows), associated open fields (black and orange arrows), the mass flow (blue arrows), and rope-like structures (white arrows in panels (d-f)). The diamond implies the position of wave onset, far away from the flare (the white arrow in panel (a)). (g) The fan-spine topology (red and green arrows) of the flare in the persistence image of AIA 94~{\AA} between 19:20 and 20:40 UT. (h) The intensity curve in AIA 94~{\AA} in the dotted box in panel (a), and the dashed line marks the beginning time of the flare.
An animation of the AIA 131 and 94~{\AA} images is available. The video begins at 19:50:15 UT and ends at 20:29:15 UT. The realtime duration of the video is 1 second.
\label{f6}}
\end{figure}

\clearpage

\begin{figure}
\epsscale{1.0} \plotone{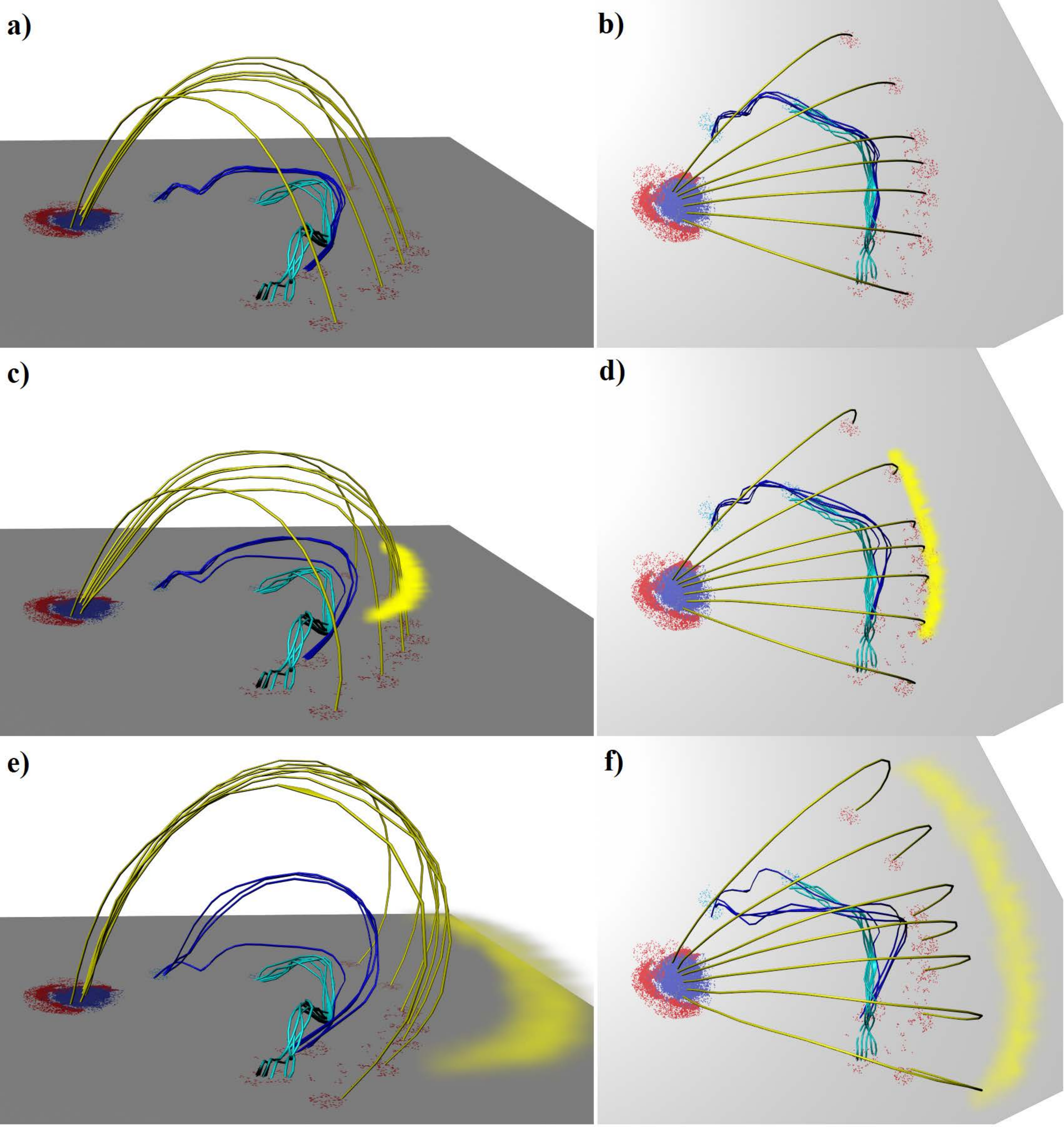}
\caption{The scenario of the event. The schematic representation in side (left panels) and top (right panels) views displays F1 (blue lines), F2 (cyan lines), the overlying loops (yellow lines), and the EUV wave (the bright yellow shade). The blue and red patches at the base imply the negative and positive polarities, respectively.
\label{f7}}
\end{figure}

\end{document}